\documentclass[pre,floats,twocolumn,aps,superscriptaddress,showpacs]{revtex4}
\usepackage{amsmath}
\usepackage{epsfig}

\newcommand{\bx}{{\bf  x}}
\newcommand{\bq}{{\bf  q}}
\newcommand{\br}{{\bf  r}}
\newcommand{\ve}{{\bf  e}}
\newcommand{\val}{{\vec \alpha}}
\newcommand{\vs}{{\vec S}}
\newcommand{\vv}{{\vec v}}

\newcommand{\BE}{\begin{equation}}
\newcommand{\EE}{\end{equation}}
\newcommand{\beqn}{\begin{eqnarray}}
\newcommand{\eeqn}{\end{eqnarray}}

\begin{document}

\title{
Clustering, advection and patterns in a model of population dynamics
with neighborhood-dependent rates.
}

\author{Emilio Hern\'andez-Garc\'\i a}
\affiliation{Instituto
Mediterr\'aneo de Estudios Avanzados IMEDEA (CSIC-UIB),
    Campus de la Universitat de les Illes Balears,
E-07122 Palma de Mallorca, Spain.
}
\author{Crist\'obal L\'opez}
\affiliation{Instituto
Mediterr\'aneo de Estudios Avanzados IMEDEA (CSIC-UIB),
    Campus de la Universitat de les Illes Balears,
E-07122 Palma de Mallorca, Spain.
}
\affiliation{
Departament de F{\'\i}sica, Universitat de les Illes
Balears, E-07122 Palma de Mallorca, Spain.
}
\date{October 16, 2003}

\begin{abstract}

      We introduce a simple model of population dynamics which considers
birth and death rates for every individual that depend on the number of
particles in its neighborhood. The model shows an inhomogeneous
quasistationary pattern with many different clusters of particles.
      We derive the equation for the macroscopic density of particles,
perform a linear stability analysis on it, and show that there is a
finite-wavelength
instability leading to pattern formation. This is the responsible for
the approximate periodicity with which the clusters of particles arrange
in the microscopic model.
      In addition, we consider the population when immersed in a fluid
medium and analyze the influence of advection on global properties of
the model.

\end{abstract}
\maketitle



\section{Introduction}
\label{sec:intro}

Interacting particle systems are useful models to understand a variety
of effects in fields
as diverse as condensed matter physics, chemical kinetics, population
biology
    (where they are called individual based models) or sociology
    (agent based models) \cite{MarroDickman}. As one of the
simplest examples one can consider an ensemble of Brownian particles,
each one dying or duplicating with given probabilities per unit of time.
Several authors 
\cite{YoungNature,Adler97,superprocesses,Zhang90,Houchmandzadeh02} have
considered such {\it Brownian Bug} (BB) model in the context of
population dynamics (in particular to address plankton distributions and
patchiness), in the case in which the probabilities of death and
reproduction are equal. Aggregation of the particles in a decreasing
number of clusters occurs. This clustering is somehow surprising since a
standard mean-field or rate-equation description gives for the particle
density $\rho$ the equation
\begin{equation}
\frac{\partial \rho}{\partial t}=(\lambda_0-\beta_0)\rho+D \nabla^2 \rho,
\label{dif}
\end{equation}
where $D$ is the diffusion coefficient, and $\lambda_0$ and $\beta_0$
are the birth and death rates, respectively. Obviously,
when $\lambda_0=\beta_0$ Eq. (\ref{dif}) is simply the diffusion equation
which cannot
lead to spatial inhomogeneities.

This result was known since some time ago for this and related
models\cite{YoungNature,Adler97,superprocesses,Zhang90,Houchmandzadeh02,Meyer96,Cardy96,Shnerb00}, 
and points out the relevance of the fluctuations present in the discrete
stochastic particle model, neglected in a na{\"\i}f mean-field
macroscopic description, and that lead to {\sl reproductive pair
correlations}: the mean rates of death and birth are equal, but if
locally there is an excess of reproduction events, and diffusion is not
fast enough, a cluster of particles will develop, whereas no birth will
occur in empty zones and particles will simply disappear from regions
with excess of death.

The authors of \cite{YoungNature} go beyond that result, and show that
the clustering
    persists even in the presence of rather strong stirring, as it would
occur if the bugs
    live in a turbulent fluid such as the Ocean. The clusters now become
elongated filaments,
    but there is still strong spatial inhomogeneity arising from the
microscopic particle
    fluctuations and reproductive correlations.

The simple model just described misses some important features present
in real biological
populations. The most obvious is the absence of any interaction between
the bugs.
    Among other consequences, the global dynamics of the system, i.e. the
time evolution
    of the total number of particles, is completely independent of its
spatial distribution.
    Thus stirring the system alters the spatial pattern of the bugs,
but neither their
    individual lifetimes, nor the time history of the particle number, nor
its statistical properties.


In this Paper we introduce interacting particle models
by modifying the birth and death rates of the BB model. They will take
into account the number of neighbors within a given distance of each
bug. There is now a
strong interplay between the bug dynamics and the ambient flow and, in
addition, new effects arising from the spatial range of interaction
occur and
modify the reproductive-correlations clustering effect. In particular,
an inhomogeneous
steady structure with many different clusters of particles coming from
different
families (i.e. they are born from a different parent),
   and arranged in a periodic pattern, may occur. The number of
particles in any of these clusters is similar, resembling the spreading
of individuals in small groups over a geographical
area. This pattern formation phenomenon occurs via a finite wavelength
instability that can be characterized in a deterministic description,
being fluctuations only of secondary importance. We analyze the
phenomenon with a continuous-field Langevin description obtained from
the particle model by Fock space techniques.

The Paper is organized as follows. In the next Section we introduce the
discrete
models. In Sect. \ref{sec:numerical} we study numerically some of their
properties.
In \ref{sec:stability} we study the pattern formation process and
perform a stability analysis within the  continuum-field description of our
model. In Section \ref{sec:flow} we study the influence of a fluid flow
on the
particle system, and in the last Section we summarize our Conclusions.

Derivation of the continuum-field representation for our particle model
is interesting by itself. In this Paper we will use (Section
\ref{sec:stability}) just the deterministic part of this representation,
but we provide a detailed derivation of the full Langevin equation, via
Fock space techniques, in the Appendix, so that the additional results
presented there can be used for future reference. Representation of
advection processes within the Fock space formalism is also considered
there.

\section{Models}
\label{sec:models}

In this section we introduce the discrete models subject of study of the
paper.
We begin defining the original BB model and then our extensions.

\subsection{BB model}

The microscopic rules are simply enumerated \cite{YoungNature}. Let
$N(t)$ the number of bugs in
the system (a two-dimensional periodic box of size $L\times L$; in all
our computer simulations we will take $L=1$):
\begin{enumerate}

\item There is an initial population of $N(t=0)=N_0$ bugs or particles,
randomly located.

\item One particle is selected at random and it dies
with probability $p$, reproduces with probability $q$, or remains unchanged
    with probability $r$ ($p+q+r=1$). In the case of reproduction, the
newborn particle is
located at the same place as the parent particle. The process is repeated a
    number  $N(t)$ of times \cite{asynchronous}.

\item Each particle moves in random direction a distance drawn
    from a Gaussian distribution of standard deviation $\sigma$ (this
models Brownian motion).

\item When advection is considered,
the particles are transported by an external flow to be described later.

\item Time is incremented an amount $\tau$, and the algorithm repeats.
\end{enumerate}

Symbolically, in chemical reaction notation:
\begin{eqnarray}
A & \stackrel{\beta_0}{\longrightarrow}& \emptyset,  \label{anhi}  \\
A & \stackrel{\lambda_0}{\longrightarrow}& A + A     \label{repro}
\end{eqnarray}
where $A$ represents individual particles, each one dying at a rate
$\beta_0=p/\tau$
(death rate per particle and unit of time), or reproducing at
a rate $\lambda_0=q/\tau$. The Brownian motion step leads to diffusion
with a diffusion coefficient $D=\sigma^2/2\tau$. In the following we
measure time in units of $\tau$, so that $\tau=1$, $\beta_0=p$,
$\lambda_0=q$, and $\sigma=\sqrt{2D}$.
We take $r=0$ so that $\lambda_0+\beta_0=1$, and define the important
parameter $\mu=\lambda_0-\beta_0$, the difference between birth and
death rates.

\subsection{Neighbourhood-dependent (ND) model}

The new model is analogous to the one before, except that in step $2$
the reproduction and death rates of a given particle labelled $j$,
    $\lambda(j)$ and $\beta(j)$, and are not constant but depend on the
number of
particles surrounding the particle $j$. Explicitly we take (with $\tau=1$):
\begin{equation}
\lambda(j)=\max\left( 0, \lambda_0-\frac{1}{N_s} N_R^j  \right) ,
\label{repdep}
\end{equation}
and
\begin{equation}
\beta(j)=\max\left( 0, \beta_0-\frac{\alpha}{N_s} N_R^j  \right)\ .
\label{deathdep}
\end{equation}
where $ N_R^j$ denotes the total number of particles which are at a distance
smaller than $R$ from particle $j$ (excluding the particle $j$ itself).
$R$ is thus a  range of interaction, $N_s$ is
a saturation parameter, and $\alpha$ controls the asymmetry between
its influence on death and on reproduction. The BB model is recovered
when $R \rightarrow 0$. The {\sl maximum} condition is imposed to
insure that $\lambda(j)$ and $\beta(j)$ are positive definite, as it
should be given that they are probabilities. When $N_s$ is positive,
the model penalizes reproduction when particles are crowdedly
surrounded. Lonely particles reproduce
with higher probability. This kind of interaction would be appropriate
to model individuals that compete for resources (e.g. food), in a
neighborhood of its actual position.
When $\frac{\alpha}{N_s}$ is positive, death rate decreases in
crowded environments, modelling a kind of mutual protection. The opposite
behavior occurs when these parameters are negative, a situation that will
not be considered in the present Paper, though most of the results
presented can be
extended easily to this case.  This model of interacting particles is
related to many others (see reviews in
\cite{MarroDickman,Hinrichsen2000}) in which some limitation in the
growth of the population at a single site is imposed via a fermionic
restriction (explicitly stated \cite{Moro2003} or implicitly imposed on
computer simulations by forbidding double occupation of lattice sites)
or via the inclusion of the coagulation process $A+A \rightarrow A$
inverse to (\ref{repro}) \cite{Cardy9698,Chate2003}. Our model (and the
BB model) shares
with them several qualitative features, being the most important the
fact that the empty state is an absorbing state: if at some moment all
the population becomes extinct, no recovery is possible within the rules
of the model. This leads to an {\sl absorbing phase transition} from an
active or surviving phase to an absorbing dead phase when some effective
reproduction rate is reduced. The peculiarity in our model is that
interactions are not purely local, but extend to a finite distance $R$.
This should be irrelevant for the critical behavior close to the
absorbing phase transition, since only asymptotically large scales are
important there,  and we expect this transition in our model to be in
the standard Directed Percolation (DP) or Reggeon field-theory
universality class to which many of these interacting models belong
\cite{Hinrichsen2000,mamunoz}. We will see, however, that the behavior
in the active phase is greatly influenced by the existence of a finite
interaction range $R$. In consequence we will not analyze in great
detail the absorbing phase transition, but concentrate in the active
phase(s), where more novel behavior occurs.


\section{Numerical study of the discrete models}
\label{sec:numerical}

This section is devoted to present some numerical results that
    stress the differences between
the BB and the ND model. Here we  consider the system with no external
driving flow,
whose analysis is left to section \ref{sec:flow}.

\begin{figure}
\epsfig{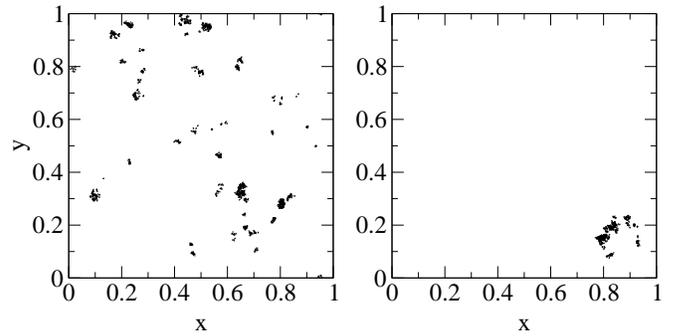}
\caption{
   Spatial configurations for the
BB model
at two different times. Left: configuration after $100$ steps, with a
large number of surviving clusters. Right correspond to a single cluster
remaining after $3000$ steps. The value of the parameters are
$\lambda_0=\beta_0=0.5$, $D=10^{-5}$, and the initial population is of
$N_0=1500$ bugs
randomly distributed.
\vspace*{0.5cm}}
\label{fig:youngpatron}
\end{figure}

The BB model has been studied in
detail~\cite{YoungNature,Adler97,Zhang90,superprocesses}.
If $\mu=\lambda_0-\beta_0>0$, the total population generally explodes
exponentially, with a time scale given by $\mu^{-1}$, although there is
a finite probability for extinction that depends on the initial
population and decreases with increasing $\mu$. If $\mu<0$ the final
state is, with probability 1, the total extinction of the particles,
occurring again at a  exponential rate characterized on average by
$\mu^{-1}$, but with diverging relative fluctuations. A critical
situation occurs when $\lambda_0=\beta_0$. In this case, the particles
arising
from the same ancestor form clusters, with the number of clusters
decreasing in
time and the number of particles in the surviving clusters growing, so that
$\left<N(t)\right>=N_0$ $\forall t$, with the average taken over
different realizations.
But fluctuations in $N(t)$
are huge (its variance diverges linearly in time), with some runs leading
to fast extinctions, and others with clusters surviving for long time.
In a finite system all clusters finally disappear, but the typical
life-time diverges
linearly with $N_0$, and the average life time is infinity
\cite{Zhang90}. Fig. \ref{fig:youngpatron} shows
the distribution of particles at two different stages of the evolution,
one in which a large single cluster, coming from a single ancestor, is 
present
    (right panel) corresponding
to a long-time evolution, and another with still many clusters from
different ancestors (left panel)
for an earlier stage of the temporal evolution.

\begin{figure}
\epsfig{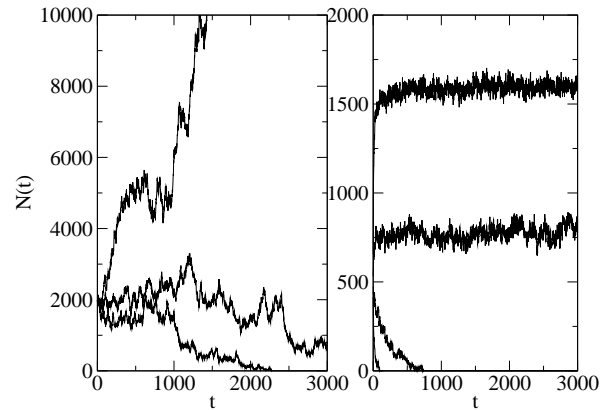}
\caption{a) Total number of particles, $N(t)$, vs time for the BB model
and three different values of the control parameter $\mu$: From top to
bottom: $\mu=5\ 10^{-4}$, $\mu=0$, and $\mu=-5\ 10^{-4}$; here
$D=10^{-5}$. b)
Idem for the ND model and four different values of $\mu$:  From top to
bottom, $\mu=0.7$, $\mu=0.5$, $\mu=0.4$, and $\mu=0.3$. Two are above
critical ($\mu_c \approx 0.4$) and two below it.
The other parameter values are $R=0.1$, $N_s=50$, and $D=10^{-5}$.
}
\label{fig:tribalbb}
\end{figure}

Figure \ref{fig:tribalbb}a)
shows the time evolution of the total number of bugs $N(t)$ in the
critical case,
$\mu=0$, for
a particular realization, displaying the critical fluctuations, and
examples of cases with nonvanishing $\mu$. One can observe the fast
decay (growth)
of $N(t)$ for  $\mu$ negative (positive).

\begin{figure}
\epsfig{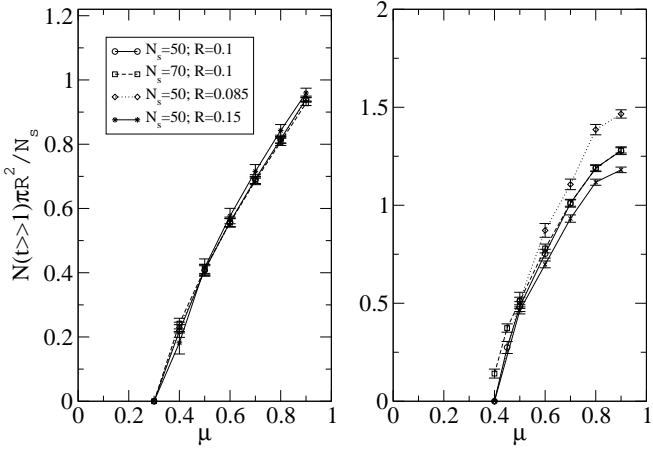}
\caption{Long-time average number of particles $N(t)$ vs $\mu$. Left
panel corresponds
to $D=10^{-4}$ and right to  $D=10^{-5}$, and the other parameters as
indicated. The average is taken from the instantaneous particle numbers
at times between $1000$ and $10000$ steps; the
error bar indicates the standard deviation of the instantaneous
fluctuations around this mean value.
}
\label{fig:bifurcation}
\end{figure}

    The behavior of the ND model is rather different. Just for simplicity
we consider here
(and in the rest of the Paper) the value $\alpha=0$ so that only
reproduction
depends on the neighborhood. Figure~\ref{fig:tribalbb}b)
   shows the time evolution of
the population. For $\mu$ smaller than a critical value $\mu_c>0$ (which
turns out to be $\mu_c \approx 0.4$ for the parameter values used in the
Figure), we find always
extinction, whereas typical realizations reach a finite average
population at long times for $\mu>\mu_c$. We plot
in Fig.~\ref{fig:bifurcation} the total average number of particles
$N(t)$ at
long-time vs $\mu$, and different values of the parameters. The scaling
used to
present $N(t)$ is suggested from an analytical expression discussed in
next Section. As discussed later, it provides good data collapse in the
left plot Fig.~\ref{fig:bifurcation}, for a diffusion coefficient of
$D=10^{-4}$, but it is
grossly inadequate for data in the right panel,
corresponding to smaller diffusivity, $D=10^{-5}$.

\begin{figure}
\epsfig{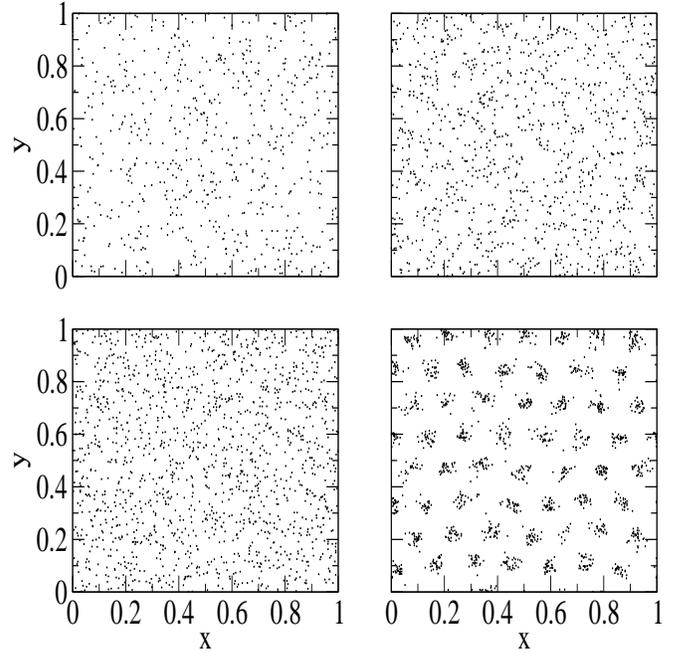}
\caption{Long-time spatial structures for the ND model.
Left column corresponds to two patterns with the same value of
$D=10^{-4}$, and two different values of $\mu=0.5$ (up) and
$\mu=0.9$ (bottom).
Right column corresponds to fixed $\mu=0.7$, and $D=10^{-4}$ (upper), and
$D=10^{-5}$  (bottom). In all the plots, $N_s=50$ and $R=0.1$.
}
\label{fig:tribal}
\end{figure}

   The nature of the
spatial distribution in the active phase depends on the values of the
parameters. For large enough $D$, the spatial distribution of particles
is homogeneous on
average, whereas clear clustering occurs for small $D$. As in the BB
model, the clusters are coming from different families. But here they
are {\it not transient} and the most striking feature is that they
{\it  organize in
a periodic pattern}. The periodicity of the pattern is of the order of
$R$, the interaction range.
   In addition to decreasing $D$, this transition to a  periodic
organization occurs by increasing $R$ and, for small enough $D$, by
increasing $\mu$. Fig.~\ref{fig:tribal}
shows examples of the different spatial patterns.

{\it We see that the most notable effect of the introduction of
interactions
with a characteristic spatial scale has been the segregation of bugs in a
periodic array of clusters} (bottom-right plot in Fig. ~\ref{fig:tribal}).
   This seem to be a rather natural way to
make compatible the high local growth at relatively large value of
$\mu$, with the reduction of this growth that a too crowded neighborhood
would imply: the empty space between the clusters acts as a buffer zone
keeping the competition for resources less limiting than in a
homogeneous distribution. We expect that this mechanism will appear in
Nature when there is a scattering of the total population in small
groups over a large spatial area. One  can think, for instance, of the
spreading of groups of predators, or even of primitive human societies
that are aggregated in small tribes.

We characterize the patterns in terms of the {\sl structure factor}
$S(k)$. It is defined as
\BE
S(K)=\left\langle  \left|\frac{1}{N(t)} \sum_j e^{i \bq \cdot \bx_j(t)}
\right|^2 \
\right\rangle _{K,t}
\label{factes}
\EE
where $\bx_j(t)=(x_j,y_j)$ is the position vector of the particle $j$ at
time $t$, $\bq=(q_x,q_y)$ is a two-dimensional wavevector, the sum is
over all particles, and the average is a spherical average over all
wavevectors of modulus $|\bq|=K$, and a further temporal average in the
long-time state is added to improve statistics. Maxima in this function
identify
relevant periodicities in the interparticle distribution.

\begin{figure}
\epsfig{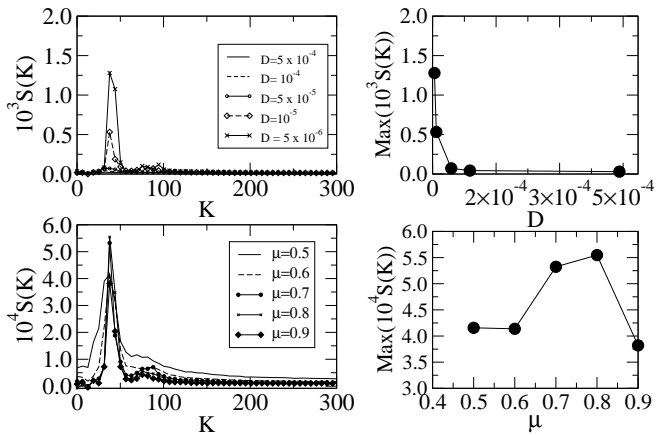}
\caption{Structure factors (left panels) and corresponding height of the
main peak (right) for different patterns in the
ND model. Upper panels are for fixed value of $\mu=0.7$ and the values
of $D$ shown in the legend box.
Bottom panels are for $D=10^{-5}$ and different $\mu$'s as shown
in the legend box. In all the plots, $N_s=50$ and $R=0.1$.
  Note the different scales in all the plots.
}
\label{fig:tribalestructura}
\end{figure}

In Fig.~\ref{fig:tribalestructura} we show the structure factor in
the steady state
of the model for different values of the parameters. The emergence of
the periodic
patterns is indicated by the peak in the structure factor.
As in Fig.~\ref{fig:tribal}, we show here two different scenarios: The
upper panels correspond to the structure factor (left) of different patterns
with $\mu$ fixed and changing $D$. One can observe that by decreasing
$D$ the
value of the peak increases (upper-right panel), indicating that clustering
with a strong periodicity develops. Bottom panels are for $D=10^{-5}$
and different
values of $\mu$. Here the pattern is rather developed at all values of
$\mu$ above the absorbing transition at $\mu_c \approx 0.4$, with only
mild variations of the peak height (right panel) with $\mu$. By
analyzing non-spherically-averaged versions of (\ref{factes}), we
confirm that the periodic pattern has hexagonal symmetry at onset,
although this symmetry changes at high values of $\mu$.

We next try to explain quantitatively the observed patterns in terms
of an analytical description.

\section{Field theory description, stability analysis, and spatial patterns}
\label{sec:stability}

Standard theories and tools for pattern formation studies
\cite{patterns} address continuous field models, and are not particulary
well suited to analyze pattern formation in particle systems.
Fortunately, there are well established techniques (known under the name
of Doi-Peliti theory or Fock space techniques
\cite{doipeliti,Cardy96,summaries}) that allow a description of
interacting particle systems in terms of field-theoretic Langevin
equations. These techniques turn out to be equivalent to the Poisson
representation \cite{Gardiner,PoissonRepresentation}.
In the simplest
cases, the interacting particles are instantaneously Poisson distributed
in each small space region, and the field description gives the space
and time varying average value $\phi(\bx,t)$ of the local Poisson
distribution for the particle density $\rho$.

In general, however, the continuous field is complex and this simple
interpretation does not hold, but still in this case
all the moments of the
particle density $\rho$ can be obtained from the moments of the
fluctuating field $\phi$. For example the first moments of both
quantities are equal
$<\rho(\bx,t)>=<\phi(\bx,t)>. $
    In the Appendix, we derive in detail the Langevin field description
for the ND model
(including hydrodynamic flow, arbitrary values of $\alpha$, etc.). Two
approximations are needed to arrive to the final form
(\ref{langevincompleta}) (see the Appendix for details). As a
first attack to the problem of pattern formation in our particle model,
we analyze in this Section just the {\it deterministic} part of the field
equation, i.e., the noise term will be neglected.
   We will see that this will be sufficient to understand the
main qualitative features of the pattern forming instability. The expected
influence of the noise would be to affect system properties in the vicinity
of transitions and instabilities, and to shift the position of the
transition lines~\cite{mamunoz}.
   In addition, since our system is translational invariant
and two-dimensional, it is very likely that the  sharp bifurcation to
patterns that we find in the deterministic analysis will be blurred by
noise into a non-sharp crossover even in the thermodynamic limit.
Nevertheless, we find good qualitative agreement between most of the
observed properties of the discrete model presented in the preceding
section  and the deterministic predictions obtained
here.

Thus we analyze the deterministic version of
Eq.~(\ref{langevincompleta}) (no flow and $\alpha=0$):
\begin{eqnarray}
&\partial_t \phi (\bx,t) =   D  \nabla^2   \phi(\bx,t)  +  \nonumber  \\
&(\lambda_0 - \beta_0) \phi(\bx,t)
- \frac{1}{N_s} \phi(\bx,t) \int_{|\bx-\br|<R}d\br \ \phi(\br,t)
\label{deterministatribal}
\end{eqnarray}
At this level of mean-field-like approximation (no noise), the field
$\phi(\bx,t)$ can be interpreted as the density field $\rho(\bx,t)$.
Stationary homogenous solutions of this equation are the absorbing phase
$\phi (\bx,t)=0$, and the {\it active} or {\it survival} phase
$\phi (\bx,t)=\phi_s=\mu N_s/\pi R^2$ (remember that   $\mu
=\lambda_0-\beta_0$).
    For $\mu <0$ the only stable solution
is the absorbing one; the  transition to the survival state is approached
at $\mu =0$, and this state is stable for a range of positive values of
$\mu$. At the deterministic
level the transition is transcritical, but we expect it to be modified
by noise into a transition of DP type \cite{mamunoz,Hinrichsen2000},
occurring at values of $\mu$ larger than zero, as observed.

\begin{figure}
\epsfig{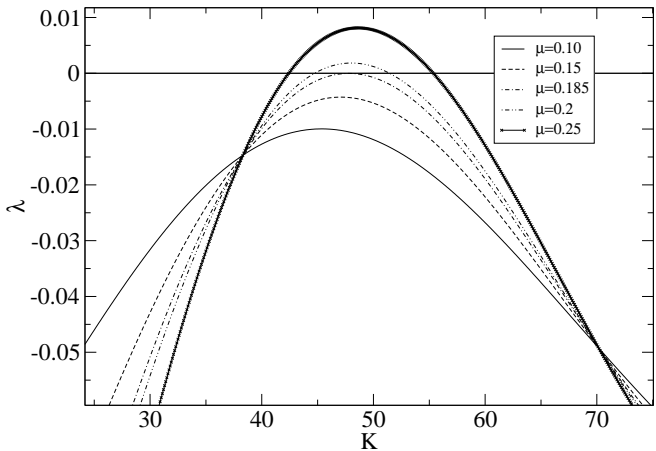}
\caption{Linear growth rate $\lambda$ vs wavenumber $K$ from
(\ref{dispersion})
for different values of $\mu$ close to $\mu_P$.
We take $R=0.1$ and  $D=10^{-5}$ so that $\mu_P=0.185$ and $K_m=47.79$.
}
\label{fig:dispersion}
\end{figure}

We make a stability analysis of the
$\phi_s$ solution by considering small harmonic perturbations  around it,
$\phi (\bx,t)=\phi_s+\delta \phi (\bx,t)$, with
$\delta \phi (\bx,t) \propto \exp(\lambda t + i {\bf k} \cdot \bx)$.
After simple calculations  one arrives to the following
dispersion relation
\begin{equation}
\lambda(K) =-D K^2 -\frac{2 \mu}{K R}J_1 (KR),
\label{dispersion}
\end{equation}
where $K$ is the modulus of ${\bf k}$, and $J_1$ is the first-order
Bessel function. It is clear that the relevant parameters in the problem
are $\mu$ and $D/R^2$ (in fact the precise adimensional combinations are
$\mu\tau$ and $D\tau/R^2$, but remember that we are measuring times in
units of $\tau$, so that $\tau=1$). The eigenvalue $\lambda(K)$ (which
is in fact a function of $KR$, $\mu$, and $D/R^2$) is real and can be
positive
for some values of the parameters. This is shown in Fig.
(\ref{fig:dispersion})
where we plot $\lambda$ against $K$ for different values of $\mu$ around
$\mu_P$ as given below in Eq. (\ref{eq:mucritico}),
with fixed $D/R^2$.

The equations
\begin{eqnarray}
\frac{\partial \lambda (K)}{\partial K}\mid_{K=K_m}&=&0 \\
\lambda (K_m)&=&0.
\label{condiciones}
\end{eqnarray}
identify the values of the parameters at which the maximum of the curve
$\lambda(K)$, at $K_m$, becomes positive. This gives a line of
instability $\mu_P=\mu_P(D/R^2)$ in the parameter plane. It is
straightforward to obtain that
$\mu_P={-D R K_m^3}/({2J_1(K_m R)})$ and the equation for $K_m$ reads
\begin{equation}
\frac{K_m R}{2 J_1 (K_m R)}(J_0(K_m R)-J_2 (K_m R))-3 =0.
\label{eq:K_m}
\end{equation}
$J_0$ and $J_2$ are the zero and second order Bessel functions,
respectively. This equation can be solved numerically to obtain
\BE
K_m \approx \frac{4.779}{R}
\label{eq:K_m2}
\EE
so that
\begin{equation}
\mu_P \approx 185.192 \frac{D}{R^2}.
\label{eq:mucritico}
\end{equation}
The behavior of the deterministic Eq.~(\ref{deterministatribal}) is now
clear: for $\mu<0$ the only stable solution is  $\phi =0$. In the
interval $0<\mu<\mu_P$ one has the homogeneous
density $\phi=\phi_s$, and for $\mu>\mu_P$ spatial patterns emerge. This
last transition can also be crossed by decreasing $D/R^2$ at fixed
$\mu>0$.

This scenario is consistent
with the results for the
particle ND model shown in Sec.~\ref{sec:models}.
In particular, note that Eq.~(\ref{eq:K_m2}) indicates that the pattern
periodicity is determined by $R$, and
is independent of other parameters of the system such as  $\mu$, $D$ and 
$N_s$.
This is in agreement with the results  for the structure factors
shown in Fig.~\ref{fig:tribalestructura}.  It is also observed there
that the  numerical
value of the dominant wavenumber in Fig.~\ref{fig:tribalestructura}  is 
close to the
predicted value   given by Eq.~(\ref{eq:K_m2}).
   Since Eq. (\ref{deterministatribal}) has no
particular symmetries, we expect on generic grounds \cite{patterns} that
hexagonal patterns would appear close to the instability. Since they
usually bifurcate subcritically we expect some range of bistability for
$\mu<\mu_P$, that may be influenced by noise. In consequence we do not
expect the transition line
(\ref{eq:mucritico}) to be fully accurate. Nevertheless it correctly
explains the distinct behavior between the data shown in
Fig.~\ref{fig:bifurcation}a (essentially all of them predicted to be in
the homogeneous phase, as confirmed by Figs.~\ref{fig:tribal} and
\ref{fig:tribalestructura}) and Fig.~\ref{fig:bifurcation}b (for smaller
$D$, so that all data points are in the periodic clustered phase). The
curves in Fig.~\ref{fig:bifurcation}a collapse together and approach the
deterministic prediction for the homogeneous solution $\phi_s$ (a
straight line of slope 1 in that scaled plot) sufficiently far from the
absorbing transition point. Such collapse does not occur in
Fig.~\ref{fig:bifurcation}b since they do not correspond to homogeneous
states. More important are the fluctuation corrections to our
deterministic results around the absorbing transition: the transition
point is quite far from the deterministic value $\mu_c=0$ and the
critical behavior is very different from the simple linear vanishing of
the number of particles predicted deterministically.

\begin{figure}
\epsfig{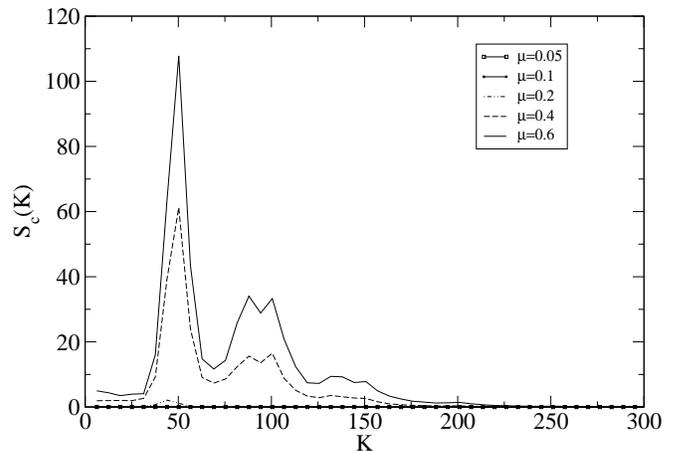}
\caption{Spherically averaged continuum structure function against $K$ for
different values of the control parameter $\mu$. The other parameter values
are $D=10^{-5}$ and $R=0.1$.
}
\label{fig:strucfun}
\end{figure}

\begin{figure}
\epsfig{figure=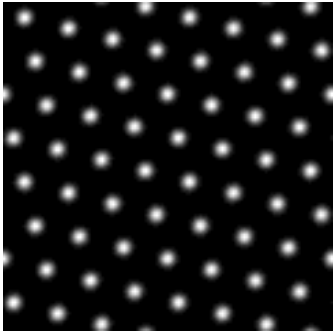,width=0.5\columnwidth,angle=0}
\caption{Steady spatial pattern from the deterministic
equation (\ref{deterministatribal}). $\mu=0.70$, $R=0.1$, $D=10^{-5}$,
and $N_s=50$.
Note the strong similarity with the pattern in the bottom-right plot in
Fig.~\ref{fig:tribal}.
}
\label{fig:patrontribal}
\end{figure}

In Fig.~\ref{fig:strucfun} we plot the spherically averaged structure
function, $S_c(K)$,
against the wavenumer, $K$,
of the density field $\phi$ obtained numerically, after a long-time, from
numerical solution of Eq.~(\ref{deterministatribal}).  $S_c(K)$ is the
modulus of the spatial Fourier transform of $\phi(\bx,t)$,
  averaged spherically and in
time. Note that, since $\phi$ is a continuous field, $S_c(K)$ is related
but not identical to the structure factor $S(K)$ of the particle
system, Eq.~(\ref{factes}). Nevertheless, maxima of $S_c(K)$ also
identify dominant periodicities.  In Fig.~\ref{fig:strucfun} we have
taken $R=0.1$, $D=10^{-5}$,
so that $\mu_P \approx 0.185$. One can see
how for $\mu>\mu_P$ the structure function develops a peak that grows
with $\mu$, indicating the development of a spatial pattern with a
typical distance between clusters. The peak
is located at the wavenumber closest to (\ref{eq:K_m2}) compatible
with the discretisation imposed by the periodic boundary conditions.
Fig.~\ref{fig:patrontribal} shows a
steady pattern of density which is analogous to the one shown for the
discrete model in
the bottom-right panel of Fig.~\ref{fig:tribal}. These observations 
confirm for the
full nonlinear model (\ref{deterministatribal}) the behavior identified
from the linear stability analysis of the homogeneous solutions.
It is also worth to mention that the non-spherically-averaged version
of the structure function displays hexagonal order for $\mu \gtrsim
\mu_P$. It degenerates into square symmetry as we increase the value of
$\mu$, indicating the appearance of additional bifurcations.

\section{Influence of fluid flow}
\label{sec:flow}

In addition to the pattern-forming instability, a crucial difference
between the BB and the neighborhood-dependent model is their
response to an external flow. Since the birth and death rates of
the BB model are fixed constants, global quantities such as the total
number of particles are independent of  any particle motion, being it
diffusive or hydrodynamic.
This seems to be rather unrealistic for applications such as modelling
plankton populations \cite{YoungNature}, always driven by external
flows. On the contrary, with the neighborhood dependence of the rates in the
tribal model we  overpass this inconvenience and the model becomes
dependent on the
environmental conditions.

\begin{figure}
\epsfig{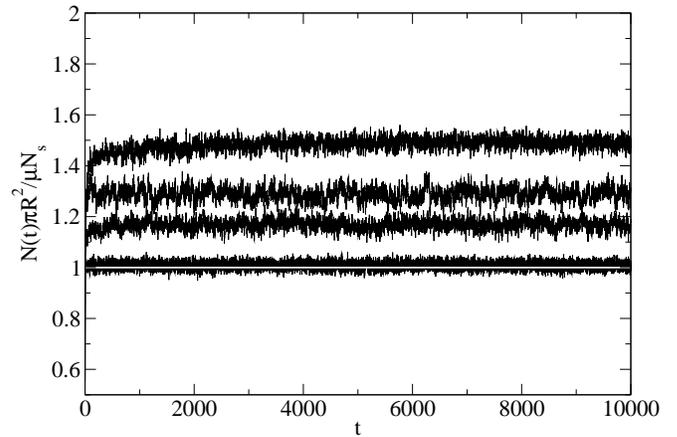}
\caption{ $N(t)$, normalized with
$\mu N_s/\pi R^2$,
vs time for different values of the external flow
strength, $A$. From top to bottom: $A=0$, $A=0.01$, $A=0.05$,
and, fluctuating around the value $1$ (white line), $A=1$ and
$A=3$. The other parameters: $\mu=0.8$, $N_s=50$, and $R=0.1$.
}
\label{fig:conflujo_08}
\end{figure}

\begin{figure}
\epsfig{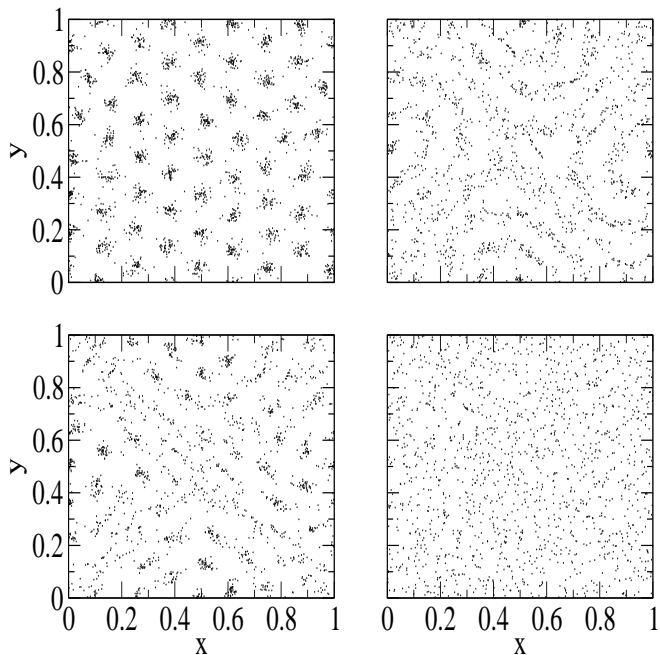}
\caption{Long-time spatial structure for the ND model with an external
flow.
  From top to bottom and left to right, $A=0$, $A=0.1$, $A=0.5$, and
$A=1$. The other parameters: $\mu=0.9$, $D=5\times 10^{-6}$, $N_s=50$, and $R=0.1$.
\vspace*{0.5cm}}
\label{fig:patternsconflujo}
\end{figure}

\begin{figure}
\epsfig{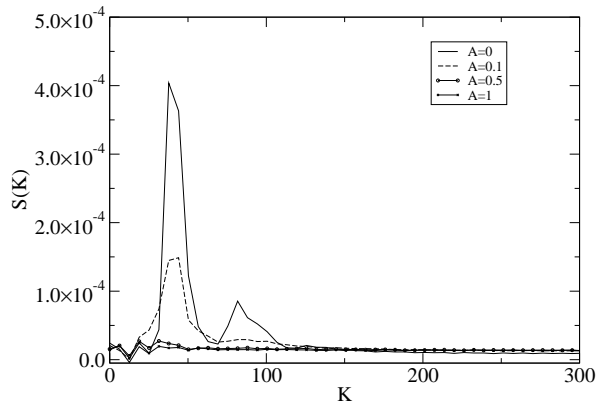}
\caption{  Structure factor for the patterns shown in
Fig.~\ref{fig:patternsconflujo}
}
\label{fig:factorestructuraflujo}
\end{figure}

As a simple illustration of the impact of a velocity field, we consider
the flow given by the Harper map, which is a symplectic map
in two dimensions and, therefore,  it resembles an incompressible flow.
   At each time step  the fourth item in
the algorithm described in Sect. \ref{sec:models} consists in moving the
particles in the following way:  If we denote by $(x_i(t),y_i(t))$  the
coordinates of the particle $i$
at time $t$, after one iteration of the map they become
\begin{eqnarray}
x_i(t')&=& x_i(t)+A \cos (y_i(t)),  \\
y_i(t')&=&y_i(t)+A \cos(x_i(t')).
\label{flow}
\end{eqnarray}
where $t'=t+\tau$. $A$ gives the strength of the flow, and depending of
its value particles can follow
regular or stochastic trajectories. Here we are just interested in
highlighting
the behavior when the flow changes, that is
when $A$ changes. The time evolution of the total number of particles is 
given in  Fig.~\ref{fig:conflujo_08}. It is
seen that the asymptotic value depends on the flow strength $A$. The
reason can be clearly seen from Fig.~\ref{fig:patternsconflujo}, where
snapshots of the particle distributions are presented. It is seen that
the periodic array of clusters in the absence of flow becomes
more filamental-like as the flow strength increases. The shape of the
filamental structures reflects the known unstable and stable foliation
of phase space for the Harper map. As for the BB model
\cite{YoungNature}, inhomogeneity persists for rather strong flow, but
finally the distribution becomes homogeneized. At this point, the
particle distribution should be very close to Poissonian, with density
given by the homogeneous solution of (\ref{deterministatribal}). This
is indeed what is observed in the figure (for large values of $A$ the
total number of particles, $N(t)$, normalized with the survival steady 
density
value, fluctuates around $1$).
  For smaller flow
strength the spatial structure in the neighborhood of each particle
becomes relevant, and the number of
particles approaches the value in the absence of flow, given in
Fig.~\ref{fig:bifurcation}. For completeness,
Fig.~\ref{fig:factorestructuraflujo} shows, in
terms of the structure factor, the disappearance of structure as the
flow strength increases.

\section{Summary}
\label{sec:summary}

In this Paper we have introduced an interacting particles model which
considers
birth and death rates for any individual depending on the number of
particles
that are within a distance smaller than a given range of interaction.
The model can be
considered as a simple non-local interacting extension of the BB model, 
but showing,
however,
a very different behavior. A striking feature is the appearence of a
quasistationary (fluctuations are present) periodic pattern of particle
clusters.

In order to deepen in the understanding of the pattern
forming instabilities and structures in
the discrete model, we have derived the continuum
equation describing the particle distribution. Two main features
characterize
this equation: a) the presence in the deterministic part of an integral
term taking into account the non-local interaction among the particles,
and b) the
very complex structure of the noise term, which also reflects the
non-locality
of the interaction. To understand the pattern forming process
we have studied the deterministic part of the equation. At this level,
the interpretation of the continuum field as the density field of
particles is valid. Then, periodic clustering appears as
a finite wavelength instability in the density equation.

In addition, we have also considered the
   response of the model to an external driving flow. Global properties
of the system like the
total number of particles depend on the flow, via the influence of it in
the spatial structure.

Future extensions of this work will include a further study
of the instabilities, an explicit consideration of the noise term in the
continuum description, and the consideration of different (chemical or
biological)
species of particles interacting through
a finite interaction range.

\begin{acknowledgments}
We acknowledge discussions with Guido Boffetta, Stefano Musacchio and
Angelo Vulpiani. We also acknowledge support from MCyT of Spain and
FEDER under projects
REN2001-0802-C02-01/MAR (IMAGEN) and BFM2000-1108 (CONOCE).
\end{acknowledgments}

\section*{APPENDIX: FIELD THEORETICAL APPROACH}
\label{sec:Appendix}

In this Appendix we derive an approximate continuum field equation
describing the dynamics of our model, by using Fock space techniques. We
present more results than needed for the analysis of Section
\ref{sec:stability}, but we include them here by its intrinsic interest
and for future reference. In addition to the usual particle interaction
and diffusion processes, we consider in this Appendix also the process
of advection, which is not usually addressed within the Fock formalism.

The standard formulation of Fock space methods \cite{Cardy96,summaries}
requires the microscopic model to be defined on a lattice. Thus we first
modify our off-lattice model to fit into a grid, apply the standard
procedure, and then let the grid spacing to become negligibly small, so
that we recover the continuum {\sl off-lattice} limit. This is somehow
different from a macroscopic limit, since only the grid size, but not
the basic lengths in our model, such as the interaction range $R$, will
be made to vanish. From the lattice approximation we can write the
stochastic dynamics in terms of a Hamiltonian operator. A path integral
representation can then be obtained and a Langevin equation can be
derived from it. The off-lattice limit is then conveniently taken.  The
whole process can not be performed exactly and we need to introduce two
approximations, one valid for not too large particle densities and
another restricting to Gaussian fluctuations.

We divide space in cells forming a lattice of $\Lambda$ sites and describe
   the state of the system by the number of particles inside each lattice
   cell $\{N_i\}_{i=1, ..., \Lambda}$. The lattice model will be equivalent
   to the off-lattice one in the limit of small lattice size, so that
only zero or one particles will be present at each site. In addition it
is convenient to work in continuous time, so that a continuous time
Markovian Master Equation describes the dynamics. At long times
($\gg\tau$) there should be no difference between the discrete and the
continuous time approaches.

Since statistically independent processes are represented by additive
terms in the Master equation, we can treat separately diffusion,
advection, and particle transformations, and then sum up the
corresponding contributions. We begin with the last process:

\subsection{Birth and death}
At each site we have a birth rate per particle
\BE
\lambda_i= \max \left( 0,\lambda_0-\frac{1}{N_s}\sum_{j \in R(i)} N_j
\right)
\label{lambdai}
\EE
and a death rate per particle
\BE
\beta_i= \max \left( 0,\beta_0-\frac{\alpha}{N_s}\sum_{j \in R(i)} N_j
\right)
\label{betai}\EE
The sum is over the cells that are within a distance $R$ from the site
$i$ (excluding cell $i$, so that in the continuum limit we recover the
original prescriptions (\ref{repdep}) and (\ref{deathdep}) ).
Defining
$P(N_1,N_2,...,N_\Lambda;t) = P(\{N \};t)$ as the probability of having
$N_1$ bugs in site $1$,...,$N_\Lambda$ bugs in site $\Lambda$ at time $t$,
we have the following Master Equation:
\begin{eqnarray}
\frac{dP(\{ N\};t)}{dt}&=&\sum_{i=1}^N (N_i-1)\lambda_i(N_i-1)
P(...,N_i-1,...;t) \nonumber \\
& +&
\sum_{i=1}^N  (N_i+1)\beta_i(N_i+1) P(...,N_i+1,...;t) \nonumber \\
& -& \sum_{i=1}^N   N_i \lambda_i(N_i) P(...,N_i,...;t) \nonumber \\
&-& \sum_{i=1}^N  N_i\beta(N_i) P(...,N_i,...;t)
\end{eqnarray}


The Fock space representation starts by defining the many-particle state
vector:
\begin{equation}
|\psi >= \sum_{N_1 ...N_\Lambda} P(\{N\};t) \prod_{i=1}^\Lambda
(a_i^\dag)^{N_i} |0>,
\label{estado}
\end{equation}
where $a_i^\dag$ and $a_i$ are the creation and annihilation operators
at lattice site $i$,
that is,
\begin{eqnarray}
a_i^\dag|N_1 ...N_i ...N_\Lambda >=|N_1 ...N_i+1 ...N_\Lambda >, \\
a_i |N_1 ...N_i ...N_\Lambda > =N_i |N_1 ...N_i-1 ...N_\Lambda >, \\
\label{accionoperadores}
\end{eqnarray}
verifying bosonic commutation rules
\begin{eqnarray}
[a_i,a_j^\dag]=\delta_{ij}, \\
a_i|0>=0.
\end{eqnarray}

Thus we have:
\begin{eqnarray}
&\frac{d|\psi>}{dt} = \sum_{\{N\}} \frac{dP(\{N\};t)}{dt}
\prod_{i=1}^\Lambda (a_i^\dag)^{N_i} |0> & \nonumber \\
&= \sum_{\{N\},i} \left[(N_i-1) \lambda_i(N_i-1) P(..N_i-1...)
\prod_{i=1}^\Lambda (a_i^\dag)^{N_i} |0>   \right.  & \nonumber   \\
&-N_i\lambda_i (N_i) P(...N_i...)\prod_{i=1}^\Lambda (a_i^\dag)^{N_i}
|0> & \nonumber \\
&+(N_i+1)\beta_i (N_i+1) P(N_i+1)\prod_{i=1}^\Lambda (a_i^\dag)^{N_i}
|0>& \nonumber \\
& \left. -N_i\beta_i(N_i)P(...N_i...)\prod_{i=1}^\Lambda 
(a_i^\dag)^{N_i} |0> \right]. &
\label{evestado}
\end{eqnarray}

After simple algebra we have a Schrodinger-like equation in Euclidean time:
\begin{equation}
\frac{d|\psi>}{dt}=-H|\psi>,
\label{schro}
\end{equation}
with the {\it Hamiltonian}:
\BE
H (a_i^\dag, a_i) = - \sum_i
\left( \left[(a_i^\dagger)^2 a_i - a_i^\dagger a_i\right] \hat \lambda_i
+ \left[a_i-a_i^\dagger a_i \right] \hat \beta_i
\right).
\label{hamiltonian}
\EE
$\hat\lambda_i$ and $\hat\beta_i$ are the operator versions of
$\lambda_i$ and $\beta_i$, that is, expressions (\ref{lambdai}) and
(\ref{betai}) with all the particle numbers $N_k$ replaced by the number
operators $a_k^\dagger a_k$. They can be defined from a power series
representation of a conveniently regularized version of the non-smooth
expressions (\ref{lambdai}) and (\ref{betai}).

To obtain a path integral representation, a first step is to use the
commutation relations in $H$ until obtaining normal ordering (i.e.,
creation operators to the left and annihilation to the right). We call
the resulting expression $H_{NO}$. Then an action depending on the
classical complex variables $\psi_i^*(t)$ and $\psi_i(t)$ is computed as:
\BE
S=\int_0^t dt \sum_i [\psi_i^* \partial_t \psi_i +H_{NO}(a_i^\dagger
\rightarrow \psi_i^*+1, a_i \rightarrow \psi_i)]
\label{fromHtoS}
\EE
   The `arrow' notation means that the operators $a_i^\dagger$ and $a_i$
should be replaced by the indicated classical variables. The action
allows to calculate average of lattice quantities such as $N_i$ as
adequate path integrals over $\psi_i$ and $\psi_i^*$ involving the
weight $e^{-S}$. For Hamiltonian (\ref{hamiltonian}), it turns out that
normal ordering leads to intricate expressions in any regularized
version of (\ref{lambdai})-(\ref{betai}). We note that for low enough
densities (or large $N_s$), the {\sl maximum} condition would be rarely
needed. In consequence, a sensible approximation at low density is:
\BE
\hat \lambda_i \approx \lambda_0-\frac{1}{N_s}\sum_{j\in R(i)}
a_j^\dagger a_j
\label{lambdapprox}\EE
and
\BE
\hat \beta_i \approx \beta_0 -\frac{\alpha}{N_s}\sum_{j\in R(i)}
a_j^\dagger a_j
\label{betapprox}\EE
In regions where densities are not small, these expressions would need
corrections. We will comment more on this later.

   The fact that $\hat\lambda_i$ and $\hat\beta_i$ do not depend on bosonic
operators on site $i$ (the sum $j\in R(i)$ excludes the cell $i$) makes
trivial the normal-ordering procedure. The action is


\begin{eqnarray}
&S=\int_0^t dt \sum_i \left\{ \psi_i^* \left[\partial_t \psi_i
+\left(\beta_0-\lambda_0 \right) \psi_i     \right. \right. &
\nonumber \\
&   \left.   +\frac{1-\alpha}{N_s} \psi_i\sum_{j \in R(i)}\psi_j  \right]&
\nonumber \\
&+(\psi_i^*)^2 \left[-\lambda_0 \psi_i + \frac{1}{N_s} \psi_i\sum_{j \in
R(i)}\psi_j  \right]   &
\nonumber \\
&   + \frac{1-\alpha}{N_s} |\psi_i|^2 \sum_{j \in R(i)}
|\psi_j|^2  &
\nonumber   \\
&+   \left.  \frac{1}{N_s} |\psi_i|^2  \psi_i^* \sum_{j \in R(i)}
|\psi_j|^2  \right\}&
\label{action}
\end{eqnarray}

An extra term (not written) should be added to implement the particular
initial condition used for $P(\{N\},t)$. One could approach the
continuum limit at this step but, for clarity,
we will perform it after writing down the Langevin equation. By
introducing Gaussian noises at each site, $\{\eta_k(t)\}$, as Gaussian
integrals in averages involving Eq.(\ref{action}), one realizes that
averages of physical quantities, such as $<N_k>$ can be obtained as
averages over stochastic processes $\{\psi_k(t)\}$ satisfying \^Ito
stochastic differential equations. Terms in the action linear in
$\{\psi^*_k\}$ give rise to deterministic terms in the stochastic
equation, terms quadratic in these variables determine the correlations
of the Gaussian noises $\{\eta_k(t)\}$. Terms of higher order give rise
to non-Gaussian noise statistics. As usual we neglect these terms (this
is our second approximation, by which we neglect the last term in
(\ref{action})) and obtain an approximate \^Ito Langevin equation:
\begin{eqnarray}
&\partial_t \psi_i (t)=\left(\lambda_0-\beta_0  \right)\psi_i &
\nonumber \\
&-\frac{1-\alpha}{N_s}\psi_i\sum_{j \in <R(i)>}\psi_j + \eta_i(t),&
\label{langevindiscreta}
\end{eqnarray}
where the Gaussian noises $\{\eta_k(t)\}$ verify:
\begin{eqnarray}
&\langle \eta_i \rangle =0 &\label{promediorumore}\\
&\langle \eta_i(t) \eta_j (t')\rangle =  2 \delta( t - t') \times  &
\nonumber \\
& \left\{ \delta_{ij}\left[\left(\lambda_0
-\frac{1}{N_s}\sum_{j\in R(i)} \psi_j \right)\psi_i  \right] \right.
&\nonumber \\
& \left.
- \theta_{ij}(R) \frac{1-\alpha}{N_s} \psi_i\psi_j \right\} .&
\label{corruido}
\end{eqnarray}
$\theta_{ij}(R)$ is a function with value $1$ if cells $i$ and $j$ are
different and each one is within a distance $R$ from the other. Noise
correlations are multiplicative, nonlocal, and rather involved. In
addition, they can not be satisfied by real stochastic process
$\{\eta_k(t)\}$ so that the noise terms, and the variables
$\{\psi_k(t)\}$, are in general complex valued. Despite this, the
particle statistics is encoded in the lattice stochastic process.  For
example, the probability for the occupation numbers is given by
\BE
P(\{N\},t)=\left\langle \prod_{i=1}^\Lambda   \frac{e^{-\psi_i(t)}
\psi_i(t)^{N_i}}{ N_i !} \right\rangle_{\{\psi_i(t)\}}
\label{Poissonaverage}\EE
where the average is over the statistics of the processes $\{\psi_i\}$.
Consequences of (\ref{Poissonaverage}) are $<N_i(t)>=<\psi_i(t)>$ and
$<N_i(t)^2>=<\psi_i(t)^2>+<\psi_i(t)>$. The term dependent on the
initial condition that was omitted from (\ref{action}) can now be taken
into account by providing adequate initial conditions for
$\{\psi_k(t)\}$, linked to the initial particle statistics by
(\ref{Poissonaverage}).

The {\sl off-lattice} continuum limit is performed by introducing the
density $\phi(\bx,t)$ by the change
$\psi_i \to \phi (\bx,t) \Delta^d$
and taking the lattice spacing going
to zero $\Delta \rightarrow 0$. $d$ is the spatial dimension ($d=2$
through this Paper). We have some freedom in performing this limit. For
example, if we take the interaction range $R$ to be a fixed number of
lattice spacings, then $R\rightarrow 0$ in the continuum limit. In this
case non-locality is lost, and by properly scaling the interaction
parameters $N_s$ and $\alpha$,
we get a continuous equation related to Reggeon field theory that has
been thoroughly studied\cite{Hinrichsen2000,mamunoz}. But we want to
describe the macroscopic behavior of the off-lattice models introduced
in Sect. \ref{sec:models}, in which the interaction range is finite.
Thus, we fix $R$ to a finite value when going to the continuum and obtain
\begin{eqnarray}
&\partial_t \phi (\bx,t)=\left( \lambda_0-\beta_0  \right) \phi(\bx,t) &
\nonumber \\
&-\frac{1-\alpha}{N_s}\phi(\bx,t) \int_{|\bx-\br|<R}d\br \phi(\br,t)
+ \eta (\bx,t),&
\label{langevincontinua}
\end{eqnarray}
with
\begin{eqnarray}
&\langle \eta (\bx,t) \rangle =0 \label{promediorumorecontinuo}&\\
&\langle \eta(\bx,t) \eta (\bx',t') \rangle= 2 \delta(t-t')  \times &
\nonumber \\
& \left\{\delta(\bx-\bx')\phi(\bx,t)
\left[\lambda_0-\frac{1}{N_s}\int_{0<|\bx-\br|<R}d\br \phi(\br,t)\right]
     \right.  &
\nonumber \\
&\left.  -\frac{1-\alpha}{N_s} \phi(\bx,t)\phi(\bx',t) \theta_{\bx
\bx'}(R) \right\}.&
\label{corruidocontinuo}
\end{eqnarray}

The continuum description of the BB model is recovered if $R \rightarrow
0$. As in the discrete case,  correlations in (\ref{corruidocontinuo})
are multiplicative, nonlocal, and lead to complex valued processes.

We see that the deterministic part of (\ref{langevincontinua}) is what
one would guess as a mean-field description for the particle density
when taking into account the finite range of the interactions. Since we
expect this to be correct for high enough densities and far from
transition points, we conclude that the terms neglected in the low
density approximation (\ref{lambdapprox})-(\ref{betapprox}) are just
correcting fluctuation statistics in a regime, high densities, in which
they are already not very relevant.

\subsection{Diffusion and advection}

Advection is implemented as a deterministic process in the off-lattice
models of Sects. (\ref{sec:models}) and (\ref{sec:flow}). When going to
a lattice description it can be implemented, for example, as cellular
automata deterministic rules. But we prefer to model it as a stochastic
birth-death process in the lattice, that in the continuum limit recovers
the deterministic flow, since this leads to a more natural description
in the Fock space formalism we are using, and the connection with the
diffusion process is clearer.

By considering stochastic hopping from cell $i$ to $j$, containing $N_i$
and $N_j$ particles, at transition rate $p(i \rightarrow j)$ per
particle, one is left with the master equation:
\begin{eqnarray}
\frac{d}{dt}P(N_i,N_j)&=&P(N_i+1,N_j-1)(N_i+1)p(i \rightarrow j)
\nonumber \\
&-&P(N_i,N_j) N_i p(i \rightarrow j) .
\end{eqnarray}
By considering the time evolution of the Fock space vector $|\psi >=
P(N_i,N_j;t) (a_i^\dag)^{N_i}(a_j^\dag)^{N_j} |0>$, one arrives to the
Hamiltonian $H=-p(i \rightarrow j)(a_j^\dagger-a_i^\dagger)a_i$. By
considering the statistically independent hopping between
an arbitrary number of cell pairs, one gets
\BE
H_{\rm hopping}=\sum_{<ij>}
(a_j^\dagger-a_i^\dagger)\left[p(j\rightarrow i)a_j-
p(i\rightarrow j) a_i\right]
\label{Hhopping}\EE
where the sum is over all different pairs of sites among which there is
hopping. In the case in which all the hopping rates are equal we recover
the standard form of the diffusion Hamiltonian
\cite{Cardy96,MattisGlasser}.

We considerer the case of a Cartesian lattice, with cells labelled by
the position $\bx$ and jumps among nearest neighbors. In this notation
(\ref{Hhopping}) becomes
\begin{eqnarray}
&H_{\rm hopping}=\frac{1}{2}\sum_\bx \sum_{\nu=1}^d \left\{
\left(   a_{\bx+\ve_\nu \Delta }^\dagger-a_{\bx}^\dagger \right) \times
     \right. \nonumber \\
&\left. \left[
p(\bx+ \ve_\nu \Delta \rightarrow \bx) a_{\bx+\ve_\nu \Delta }-p(\bx
\rightarrow \bx+ \ve_\nu \Delta ) a_\bx \right]  \right\}.
\label{Hhopping2}
\end{eqnarray}
$\{\ve_\nu\}_{\nu=1,...,d}$ are unit vectors along the positive
directions of the cartesian axes. The sum in Eq. (\ref{Hhopping2}) takes
into account all pairs of nearest neighbors. We split the jump rates
into a constant part and the local anisotropies (which will give rise to
the macroscopic flow):
\begin{eqnarray}
p(\bx \rightarrow \bx+ \ve_\nu \Delta ) &\equiv& \kappa + \alpha_\nu^+(\bx)
    \\
p(\bx+ \ve_\nu \Delta \rightarrow \bx) &\equiv& \kappa + \alpha_\nu^-(\bx +
\ve_\nu \Delta)
\end{eqnarray}
We now perform the continuum limit by letting the lattice space
$\Delta\rightarrow 0$. To this end we expand the functions
$\alpha_i^\pm(\bx + \ve_i \Delta)$ and the operators
$a^\dagger_{\bx+\ve_i \Delta }$ and $a_{\bx+\ve_i \Delta }$ in powers of
$\Delta$ to get
\begin{eqnarray}
&H_{\rm hopping} =  \nonumber  \\
&\sum_\bx  \left\{
- \Delta \nabla a_\bx^\dagger \cdot \left( \val^+(\bx)-\val^-(\bx)
\right) a_\bx  + \Delta^2 \kappa \nabla a_\bx^\dagger \cdot \nabla a_\bx
\right.  \nonumber \\
&+ \frac{\Delta^2}{2} \sum_{\nu=1}^d  \nabla_\nu a_\bx^\dagger \
\nabla_\nu \left[
\left(\alpha_\nu^+(\bx)+\alpha_\nu^-(\bx)\right)a_\bx\right]  \nonumber
\\  & \left. + {\cal O}(\Delta^3)  \right\}  .
\label{Hhopping3}
\end{eqnarray}
$\val^\pm$ are vectors of components $\alpha_\nu^\pm$. We define the
quantities $D=\kappa\Delta^2$,
$\vv(\bx)=\Delta\left(\val^+(\bx)-\val^-(\bx)\right)$, and
$\vs(\bx)=\Delta^2\left(\val^+(\bx)+\val^-(\bx)\right)/2$, and impose
them to be finite in the $\Delta \to 0$ limit. We apply the recipe
(\ref{langevincompleta}) by introducing the classical continuum fields
$a \to \phi(\bx)\Delta^d$ and $a^\dagger \to \phi^*(\bx)+1$ (Hamiltonian
(\ref{Hhopping3}) is already in normal order). In the limit $\Delta \to
0$ we find the following expression that should be added to the
continuum limit of (\ref{action}) to find the complete action of the model:
\begin{eqnarray}
&\delta S_{\rm hopping} =  \nonumber  \\
&\int dt \int d\bx  \left\{
- \nabla \phi^*(\bx) \cdot \vv(\bx) \phi(\bx)  + D \nabla \phi^*(\bx)
\cdot \nabla \phi(\bx)  \right.  \nonumber \\
& \left. + \sum_{\nu=1}^d \  \nabla_\nu \phi^*(\bx) \nabla_\nu \left(
\vs_\nu(\bx) \phi(\bx) \right) \right\}  .
\label{ShoppingFinal}
\end{eqnarray}
The Langevin representation can now be obtained. All the terms in
(\ref{ShoppingFinal}) are of first order in the variables $\phi^*$, so
that diffusion and advection only enter into the deterministic part of
the Langevin equation, not in the noise correlations.
In the simpler case in which $S(\bx)=0$ (for example this happens if
$\alpha_\nu^+(\bx)=-\alpha_\nu^-(\bx)$ or if
$\alpha_\nu^+(\bx)+\alpha_\nu^-(\bx)$ does not diverge fast enough in
the $\Delta \to 0$ limit) the complete Langevin equation reads:
\begin{eqnarray}
&\partial_t \phi (\bx,t)=-\nabla \cdot \left( \vv(\bx,t)\phi(\bx,t)
\right) + D \nabla^2 \phi(\bx,t) \nonumber \\
&+\left( \lambda_0-\beta_0  \right) \phi(\bx,t) &
\nonumber \\
&-\frac{1-\alpha}{N_s}\phi(\bx,t) \int_{|\bx-\br|<R}d\br \phi(\br,t)
+ \eta (\bx,t),&
\label{langevincompleta}
\end{eqnarray}
which allows the identification of $\vv(\bx,t)$ with a macroscopic
velocity flow (we have allowed for an explicit time dependence since
this does not alter the above derivation). The form in which the
velocity field appears in (\ref{langevincompleta}) could be guessed from
mean-field arguments. The important point of the derivation is that it
confirms that there are no extra terms in the noise correlation arising
from advection nor diffusion. The noise correlations are again given by
(\ref{promediorumorecontinuo})-(\ref{corruidocontinuo}).


\begin{thebibliography}{99}
\bibitem{asynchronous} This is asynchronous updating. In this simple BB
model, the order of the updating (synchronous, asynchronous, sequential)
is irrelevant, but it is important in the neighborhood dependent models.
\end{thebibliography}
\end{document}